\documentclass[a4paper]{jpconf}

\usepackage[square,sort&compress,numbers]{natbib}

\usepackage{graphicx}
\usepackage{amsmath,amssymb,mathrsfs}
\usepackage{wrapfig}

\newcommand{\SP}[1]{\begin{equation}\begin{split} #1
\end{split}\end{equation}}
\newcommand{\EQ}[1]{\begin{equation}\begin{split} #1
\end{split}\end{equation}}
\newcommand{\lapprox}{\raisebox{-0.5ex}{$\
\stackrel{\textstyle<}{\textstyle\sim}\ $}}

\begin{document}
\title{Overview of large $N$ QCD with chemical potential at weak and strong coupling}

\author{Timothy J. Hollowood$^1$ and Joyce C. Myers$^{2,3}$}

\address{$^1$College of Science, Swansea University, Singleton Park, Swansea SA2~8PP, UK\\$^2$Centre for Theoretical Physics, University of Groningen, Nijenborgh 4, 9747 AG Groningen, The Netherlands\\
$^3$Discovery Center, Niels Bohr Institute, University of Copenhagen, Blegdamsvej 17, 2100 Copenhagen \O, Denmark}

\ead{t.hollowood@swansea.ac.uk, jcmyers@nbi.dk}

\begin{abstract}
In this note we summarize the results from a longer article on obtaining the QCD phase diagram as a function of the temperature and chemical potential at large $N_c$ and large $N_f$ in the weak coupling limit $\lambda \rightarrow 0$, and the strong coupling limit $\lambda \rightarrow \infty$. The weak coupling phase diagram is obtained from the Polyakov line order parameter, and the quark number, calculated using $1$-loop perturbation theory for QCD formulated on $S^1 \times S^3$. The strong coupling phase diagram is obtained from the same observables calculated at leading order in the lattice strong coupling and hopping parameter expansions. We show that the matrix models in these two limits agree at temperatures and chemical potentials which are not too high, such that observables in the strongly-coupled theory can be obtained from the observables in the weakly-coupled theory, and vice versa, using a simple transformation of variables.
\end{abstract}

\section{Introduction}

QCD at non-zero chemical potential provides a description of systems at large densities, yet it is a description which is currently not directly accessible since it occurs at strong coupling and the non-zero chemical potential leads to a complex action, giving rise to the well-known sign problem. What this means is that the conventional techniques of studying finite temperature QCD: conventional lattice simulations and ordinary perturbation theory, are not applicable. In this note we consider two nonconventional perturbative techniques which allow us to calculate the partition function and related observables such as the Polyakov line and quark number, and from these to map out the phase diagram in an otherwise inaccessible range of temperatures and chemical potentials. This is an executive summary of our longer paper \cite{Hollowood:2012nr}.

The two perturbative techniques we employ are 1) weakly-coupled QCD from continuum 1-loop perturbation theory on $S^1 \times S^3$, and 2) strongly coupled lattice QCD with heavy quarks at leading order in a strong coupling, and hopping parameter expansion. In both cases we perform the calculations in the Veneziano large $N_c$ limit, where $N_c$, $N_f \rightarrow \infty$, while $\frac{N_f}{N_c}$ remains fixed. This leads to advantages in both cases (for recent reviews on the progress towards understanding gauge theories and large $N_c$ see \cite{Lucini:2012gg,Ogilvie:2012is}). In the weakly coupled theory on $S^1 \times S^3$, the large $N_c$ limit is required to have sharp phase transitions, since the calculation is only valid in very small volumes, such that $R \ll \Lambda_{QCD}^{-1}$, where $R$ is the radius of the $S^3$. For the lattice strong coupling expansion, the spatial volume is large, but large $N_c$ factorization and translational invariance lead to a simplification of the action in that terms which include correlations between different lattice sites drop out. In both cases the large $N_c$ limit allows for a description of the theory in terms of the distribution of the Polyakov line eigenvalues such that the theory reduces to an analytically solvable matrix model. For temperatures and chemical potentials which are not too high we will show that there is an exact correspondence of the matrix models of the weakly-coupled and strongly-coupled theory, under a simple change of parameters.

\section{QCD on $S^1 \times S^3$ with $\lambda \rightarrow 0$ vs. lattice QCD with heavy quarks as $\lambda \rightarrow \infty$}

The action for continuum $1$-loop QCD with constant $A_0$ was derived in \cite{Aharony:2003sx} for theories with a matter content of scalars, vectors, and/or fermions. For QCD with $N_f$ quarks of mass $m$, chemical potential $\mu$, and at temperature $T = \frac{1}{\beta}$, the action in terms of the Polyakov line observable $\rho_n \equiv \frac{1}{N_c} \Tr {\mathscr P} e^{n \int_{0}^{\beta} {\rm d}t A_0(x)} = \frac{1}{N_c} \sum_{i=1}^{N_c} e^{i n \theta_i}$ takes the form \cite{Aharony:2003sx,Hands:2010zp}
\SP{
S_{S^1\times S^3} - S_{Vdm}  = &-N_c^2 \sum_{n=1}^{\infty} \frac{1}{n} {\boldsymbol z}_{vn} \rho_n \rho_{-n}\\
&+ N_f N_c \sum_{n=1}^{\infty} \frac{(- 1)^n}{n} {\boldsymbol z}_{fn} \left( e^{n \beta \mu} \rho_n + 
e^{-n \beta \mu} \rho_{-n} \right) \, ,
\label{weak-action}
}
where $S_{Vdm}$ is the contribution from the Vandermonde determinant, and ${\boldsymbol z}_{vn}$, ${\boldsymbol z}_{fn}$ refer to the single particle partition functions for vectors and fermions,
\EQ{
{\boldsymbol z}_{vn} = 2 \sum_{l=1}^{\infty} l(l+2) e^{-n \beta (l+1)/R} = \frac{2 e^{-2 n \beta / R} (3-e^{-n \beta/R})}{(1-e^{-n \beta/R})^3} \, ,
}
\EQ{
{\boldsymbol z}_{fn} = 2 \sum_{l=1}^{\infty} l (l+1) e^{-n \frac{\beta}{R} \sqrt{(l+\frac{1}{2})^2 + m^2 R^2}} \, .
}
The action for large $N_c$, large $N_f$ lattice QCD in terms of the Polyakov line $W(x) = {\rm \Tr} \prod_{t=0}^{N_{\tau} - 1} U_{t,i}$, at leading order in the strong coupling and hopping parameter expansion is given by \cite{Damgaard:1986mx,Christensen:2012km}
\SP{
S_{lat} - S_{Vdm} = &- J D \sum_{x} \left[ \langle W \rangle W^{\dagger}(x) + \langle W^{\dagger} \rangle W(x) - \langle W \rangle \langle W^{\dagger} \rangle \right]\\
&- h N_c \sum_{x} \left[ e^{\mu \beta} W(x) + e^{-\mu \beta} W^{\dagger}(x) \right] \, ,
\label{strong-action}
}
where $J \equiv 2 \left( \frac{\beta_{lat}}{2 N_c^2} \right)^{N_{\tau}}$ for inverse coupling $\beta_{lat} = \frac{2 N_c}{g^2}$ and number of temporal slices $N_{\tau}$, and $h \equiv 2 \frac{N_f}{N_c} \kappa^{N_{\tau}}$ is the hopping parameter with $\kappa \equiv \frac{1}{a m+1+D}$ for lattice spacing $a$ and number of spatial dimensions $D$.

The actions in (\ref{weak-action}) and (\ref{strong-action}) appear fairly similar with the exception of the sum over $n$ in the former, and the sum over $x$ in the latter (the term $\langle W \rangle W^{\dagger}(x) + \langle W^{\dagger} \rangle W(x) - \langle W \rangle \langle W^{\dagger} \rangle$ in (\ref{strong-action}) compared to $\rho_1 \rho_{-1}$ in (\ref{weak-action}) leads to the same equations of motion). The lack of terms with correlations between different lattice sites in (\ref{strong-action}) means that observables of the form $\langle F(W,W^{\dagger}) \rangle$ will undergo large cancellations such that
\SP{
\langle F(W,W^{\dagger}) \rangle &= \frac{1}{N_x Z} \int \prod_x {\rm d}W(x) e^{-S[W(x),W^{\dagger}(x)]} \sum_{x'} F[W(x'),W^{\dagger}(x')] \, ,\\
&= \frac{\int {\rm d}W e^{-S(W,W^{\dagger})} F(W,W^{\dagger})}{\int {\rm d}W e^{-S(W,W^{\dagger})}} \, .
}
Therefore, when it is possible to truncate the sum over $n$ in (\ref{weak-action}) to the $n=1$ term there is an exact correspondence of matrix models resulting from (\ref{weak-action}) and (\ref{strong-action}) under the transformations
\SP{
\rho_1 &\leftrightarrow \frac{1}{N_c} \langle W \rangle\, ,\\
\rho_{-1} &\leftrightarrow \frac{1}{N_c} \langle W^{\dagger} \rangle \, ,\\
{\boldsymbol z}_{v1} &\leftrightarrow J D \, ,\\
{\boldsymbol z}_{f1} \frac{N_f}{N_c} &\leftrightarrow h \, .
}
Truncation to the $n=1$ contribution in (\ref{weak-action}) is valid when the temperature is not too high (${\boldsymbol z}_{v1}$, ${\boldsymbol z}_{f1} e^{\mu \beta} \gg {\boldsymbol z}_{v2}$, ${\boldsymbol z}_{f2} e^{2 \mu \beta}$), and the chemical potential is not too high ($\mu \lapprox \varepsilon_{f1} = \frac{1}{R} \sqrt{(l+\frac{1}{2})^2 + m^2 R^2} \big{|}_{l=1, m R \rightarrow 0}$). This region of validity includes the line of transitions extending from the temperature-axis to the chemical potential-axis, but corrections from ${\boldsymbol z}_{fn}$ for $n > 1$ would need to be included to go to higher chemical potentials, as done in \cite{Hands:2010zp}.

\section{Large $N_c$ formalism}

\begin{wrapfigure}{r}{10cm}
\includegraphics[width=10cm]{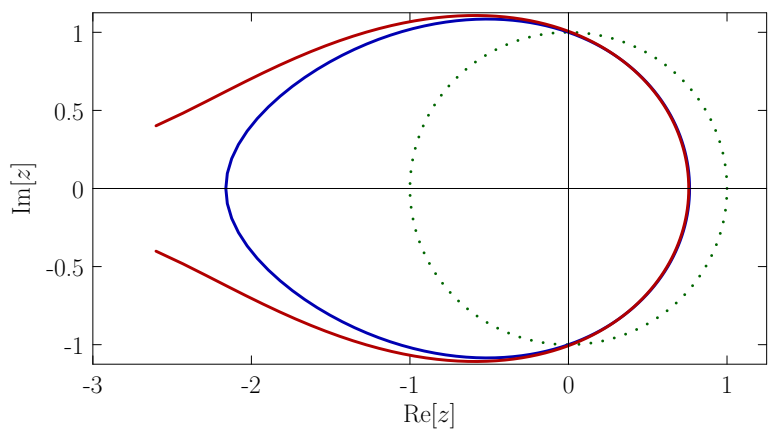}
\caption{Distribution $z(s)$ of the Polyakov line eigenvalues in the confined phase (blue, $\mu R = 0.74$), and the deconfined phase (red, $\mu R = 0.76$), for $T R = 0.3$, $\frac{N_f}{N_c} = 1$, $m R = 0$.}
\label{poly-dist}
\end{wrapfigure}

To calculate observables in the large $N_c$ limit we adopt the methods in \cite{Hands:2010zp}, which adapts the Gross-Witten-Wadia \cite{Gross:1980he,Wadia:1979vk} formalism to handle theories with complex actions. Consider a contour ${\cal C}$ with the Polyakov line eigenvalues $z_j = e^{i \theta_j}$ distributed along it with density $\varrho(z)$ defined according to the map
\EQ{
\frac{1}{N_c} \sum_{i=1}^{N_c} \xrightarrow[N_c \rightarrow \infty]{} \int_{-\psi}^{\psi} \frac{{\rm d}s}{2 \pi} = \int_{{\cal C}} \frac{{\rm d}z}{2 \pi i} \varrho(z) \, ,
}
where $\psi = \pi$ in the confined phase, such that the contour is closed, and $\psi < \pi$ in the deconfined phase, such that the contour is open and the distribution of eigenvalues $z(s)$ has a gap. This is illustrated in Figure \ref{poly-dist} as the deconfinement transition is crossed for $\mu \ne 0$.

The blue curve is a distribution $z(s)$, obtained from $i \int {\rm d}s = \int {\rm d}z \varrho(z)$, where $\varrho(z)$ is obtained in the confined phase at a point close to the deconfinement transition. Notice that the eigenvalue distribution lies significantly away from the unit circle (green dotted curve) in the $- z$ direction. The red curve is the distribution $z(s)$ obtained from $\varrho(z)$ in the deconfined (gapped) phase, at a point just past the deconfinement transition. These distributions correspond to configurations with complexified gauge fields (the $\theta_j$ are complex), which is necessary to obtain the correct stationary solution since the action is complex.

Another requirement for obtaining the correct saddle point solutions is applying the $SU(N_c)$ constraint. Without this constraint taking $\mu \ne 0$ trivially shifts $A_0$ by a constant such that the free energy is independent of $\mu$ and the quark number is always zero. The $SU(N_c)$ constraint is incorporated by adding an appropriate term to the action with a Lagrange multiplier ${\cal N}$,
\EQ{
S \rightarrow S + i {\cal N} N_c \sum_{i=1}^{N_c} \theta_i \, ,
\label{sun-term}
}
where ${\cal N} = \frac{1}{N_c^2} N_q$ is the effective quark number in the large $N_c$ limit \cite{Hands:2010zp}. The density $\varrho(z)$ is obtained by solving the equation of motion from $\frac{\partial S}{\partial \theta_i} = 0$, which, using (\ref{sun-term}) with (\ref{weak-action}) or (\ref{strong-action}) takes the form
\EQ{
{\mathfrak P} \int_{{\cal C}} \frac{{\rm d}z'}{2 \pi i} \varrho(z') \frac{z'+z}{z'-z} = \alpha_{-1} z - \alpha_1 z^{-1} -{\cal N} \, ,
\label{eom}
}
where ${\mathfrak P}$ indicates principal value and $\alpha_{\pm 1} \equiv {\boldsymbol z}_{v1} \rho_{\pm 1} + \frac{N_f}{N_c} {\boldsymbol z}_{f1} e^{\mp \mu \beta}$. The procedure for solving the equation of motion depends on whether the contour ${\cal C}$ is open, as in the deconfined phase, or closed, as in the confined phase.

\subsection{Confined (ungapped) phase}

When ${\cal C}$ is considered to be a closed contour, as in the confined phase, the equation of motion (\ref{eom}) can be solved for $\varrho(z)$ using Cauchy's theorem. In the confined phase it is sufficient to consider the Fourier expansion of the density
\EQ{
\varrho(z) = \sum_{n = -\infty}^{\infty} \rho_n z^{-n-1} \, ,
}
and solve for the $\rho_n$. The density is constrained to satisfy the identity constraint
\EQ{
\frac{1}{N_c} \sum_{i=1}^{N_c} \xrightarrow[N_c \rightarrow \infty]{} \int_{{\cal C}} \frac{{\rm d}z}{2 \pi i} \varrho(z) = 1 \, ,
}
as well as the $SU(N_c)$ constraint
\EQ{
\sum_{i=1}^{N_c} \theta_i = 0 \xrightarrow[N_c \rightarrow \infty]{} \int_{{\cal C}} \frac{{\rm d}z}{2 \pi i} \varrho(z) \log(z) = 0 \, .
}
The free energy and other relevant observables can then be calculated by plugging in the stationary point solutions obtained for the $\rho_n$.
\subsection{Deconfined (gapped) phase}

When ${\cal C}$ is considered to be an open arc, as in the deconfined phase, then the equation of motion (\ref{eom}) must be solved by defining a resolvent and solving the Plemelj formulae. The resolvent is defined from the singular integral contribution in the equation of motion
\EQ{
\phi (z) = \int_{{\cal C}} \frac{{\rm d} z'}{2 \pi i} \varrho (z') \frac{z'+z}{z'-z} \, .
}
Following \cite{Gross:1980he,Wadia:1979vk} the contour ${\cal C}$ along which the eigenvalues are distributed is defined as a square root branch cut. The resolvent can then be evaluated using singular integral techniques for open arc contours to obtain \cite{Hollowood:2012nr}
\EQ{
\phi (z) = \left( \alpha_{-1} z^1 - \alpha_1 z^{-1} \right) + \sqrt{z^2 + r^2 - 2 r x z} \left( \alpha_{1} r^{-1} z^{-1} + \alpha_{-1} \right) \, ,
}
where the endpoints of the arc occur at radius $r$ and angle $\pm \psi$, with $x \equiv \cos \psi$. Observables can then be calculated using
\[
\int_{{\cal C}} \frac{{\rm d}z}{2 \pi i} \varrho(z) F(z) = \oint_{\Gamma} \frac{{\rm d}z}{4 \pi i z} \phi(z) F(z) \, .
\]
where $\Gamma$ is defined as a contour around ${\cal C}$, which can then be peeled off to surround the residues outside such that Cauchy's theorem can be used.

\section{Results}

\begin{figure}[h]
\begin{minipage}{7.5cm}
\includegraphics[width=7.5cm]{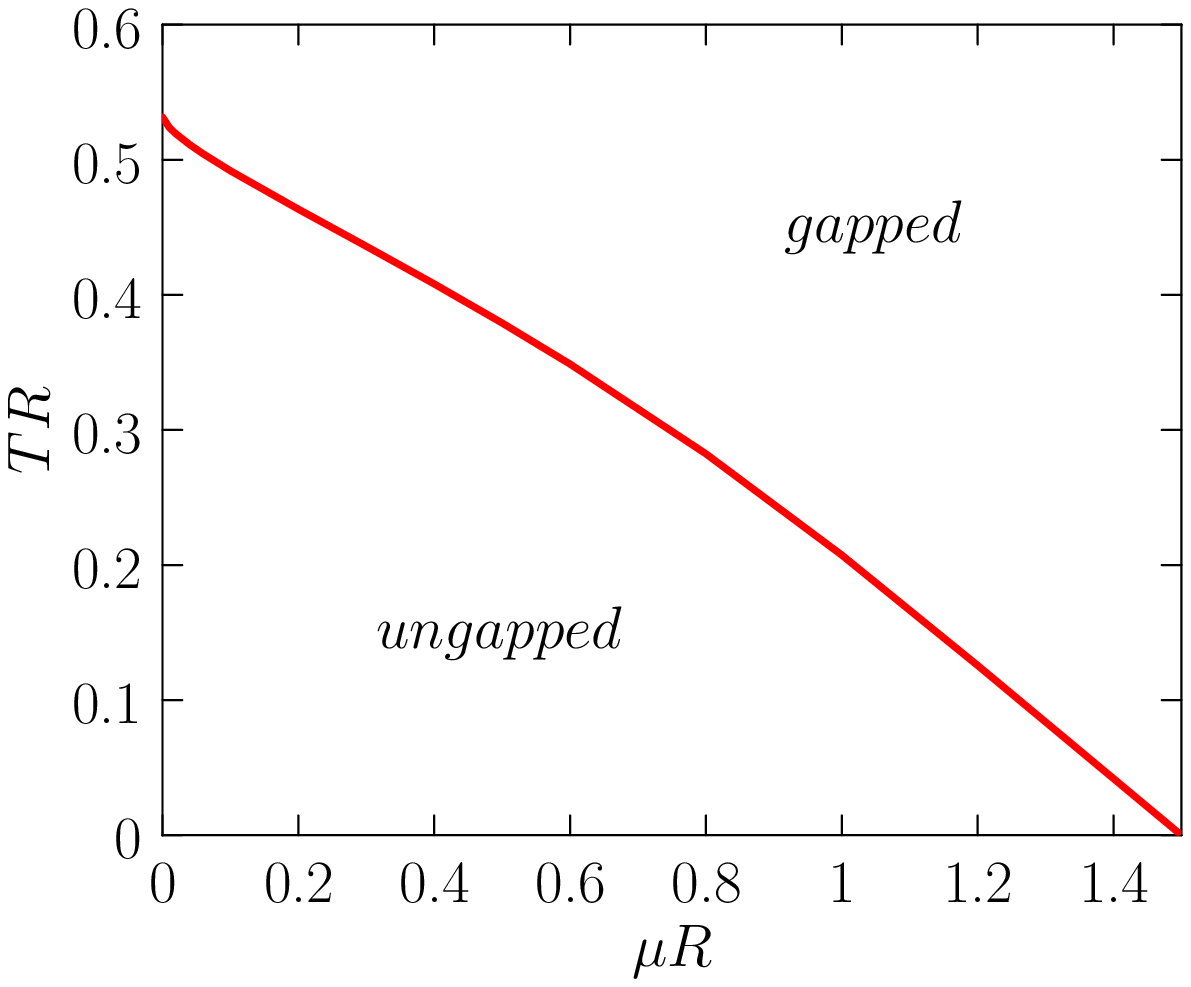}
\caption{Phase diagram on $S^1 \times S^3$ for $m R = 0$, $\frac{N_f}{N_c} = 1$.}
\label{phase-diag-weak}
\end{minipage}\hspace{2pc}%
\begin{minipage}{7.5cm}
\includegraphics[width=7.5cm]{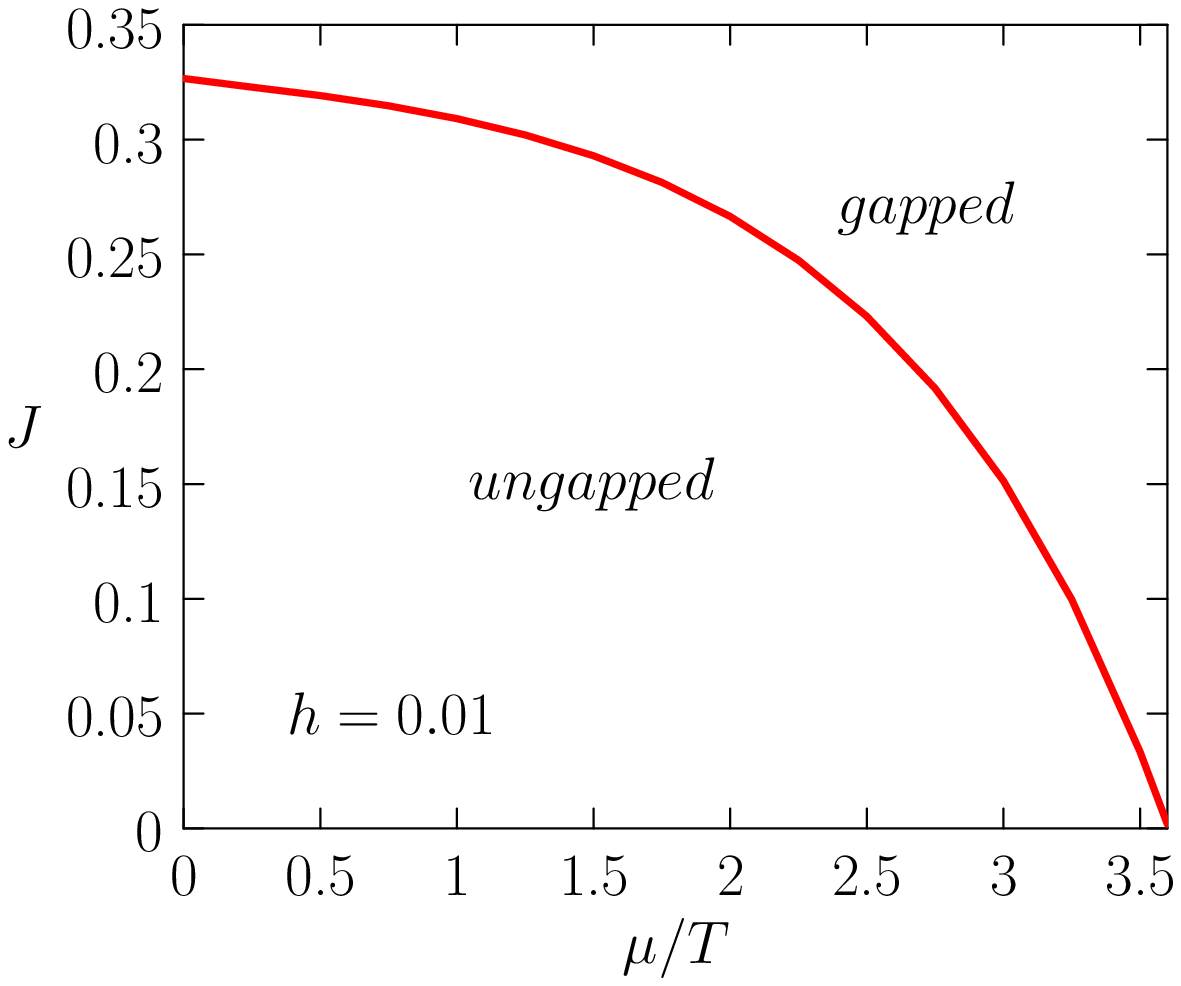}
\caption{Phase diagram from the strong coupling expansion for $h = 0.01$.}
\label{phase-diag-strong}
\end{minipage} 
\end{figure}

Using the methods of the previous sections we calculated the Polyakov lines in the confined and deconfined regions and mapped out the phase diagrams \cite{Hollowood:2012nr}. Figure \ref{phase-diag-weak} shows the phase diagram for the weakly-coupled theory and Figure \ref{phase-diag-strong} for the strongly-coupled theory. The phase boundaries are determined by comparing the free energies of the gapped and ungapped eigenvalue distributions in the regions where both are possible. In the ungapped region the effective quark number ${\cal N} = 0$ and where the gapped distribution is favored ${\cal N} \ne 0$. In both phase diagrams the order of the transition is at least fifth order at $\mu = 0$, and at least 2nd order for $\mu \ne 0$.

\bibliography{iopart-num}

\providecommand{\newblock}{}
\begin{thebibliography}{1}
\expandafter\ifx\csname url\endcsname\relax
  \def\url#1{{\tt #1}}\fi
\expandafter\ifx\csname urlprefix\endcsname\relax\def\urlprefix{URL }\fi
\providecommand{\eprint}[2][]{\url{#2}}

\bibitem{Hollowood:2012nr}
Hollowood T~J and Myers J~C 2012 {\em JHEP\/} {\bf 1210} 067 (\textit{Preprint}
  \eprint{1207.4605})

\bibitem{Lucini:2012gg}
Lucini B and Panero M 2012  (\textit{Preprint} \eprint{1210.4997})

\bibitem{Ogilvie:2012is}
Ogilvie M~C 2012 {\em J.Phys.\/} {\bf A45} 483001 (\textit{Preprint}
  \eprint{1211.2843})

\bibitem{Aharony:2003sx}
Aharony O, Marsano J, Minwalla S, Papadodimas K and Van~Raamsdonk M 2004 {\em
  Adv.Theor.Math.Phys.\/} {\bf 8} 603--696 (\textit{Preprint}
  \eprint{hep-th/0310285})

\bibitem{Hands:2010zp}
Hands S, Hollowood T~J and Myers J~C 2010 {\em JHEP\/} {\bf 1007} 086
  (\textit{Preprint} \eprint{1003.5813})

\bibitem{Damgaard:1986mx}
Damgaard P and Patkos A 1986 {\em Phys.Lett.\/} {\bf B172} 369

\bibitem{Christensen:2012km}
Christensen C~H 2012 {\em Phys.Lett.\/} {\bf B714} 306--308 (\textit{Preprint}
  \eprint{1204.2466})

\bibitem{Gross:1980he}
Gross D and Witten E 1980 {\em Phys.Rev.\/} {\bf D21} 446--453

\bibitem{Wadia:1979vk}
Wadia S 1979

\end{thebibliography}

  

\end{document}